\documentclass[twocolumn,aps,prl,superscriptaddress,floatfix,a4paper,showkeys,10pt]{revtex4-2}

\usepackage[T1]{fontenc}
\usepackage[utf8]{inputenc}
\usepackage{amsmath}
\usepackage{amssymb,bbold}
\usepackage{mathtools}
\makeatletter
\renewcommand{\p@subsection}{}
\renewcommand{\p@subsubsection}{}
\makeatother

\usepackage{xcolor}

\begin{document}
    %%%
	
	%%%%%General information
	
	\title{Doubling the magnetorheological effect of magnetic elastomers}
	
	%%% Authors	
	\author{Lukas Fischer}
	\email{lukas.fischer@ovgu.de}
	\affiliation{Institut f{\"u}r Physik, 
		Otto-von-Guericke-Universität Magdeburg, Universitätsplatz 2, 39106 Magdeburg, Germany}
	
	\author{Andreas M. Menzel}
	\email{a.menzel@ovgu.de}
	\affiliation{Institut f{\"u}r Physik, 
		Otto-von-Guericke-Universität Magdeburg, Universitätsplatz 2, 39106 Magdeburg, Germany}
	
	%%% Date
	\date{\today}
	
\begin{abstract}
    One of the most important properties of soft functionalized magnetic composite materials in view of their technological potential is given by the magnetorheological effect. It describes the change in rheological properties such as the shear modulus by application of external magnetic fields. We demonstrate how computational material design can support in approximately doubling the magnitude of this important phenomenon for magnetic elastomers. Key is to work with two perpendicular magnetic field directions. We expect future practical relevance of our concept. 
\end{abstract}

%\keywords{
%}

\maketitle

%\section{Introduction}
%\label{sec:introduction}

Combining mechanical softness with functionality is intrinsic to life. We ourselves consist of soft matter that, during evolution, developed all kinds of different functional forms. It is therefore not surprising that recent research focuses on functionalizing soft materials to mimic living matter, or at least to establish corresponding compatibility \cite{trivedi2008soft,yao2024multimodal,cao20213d}. Examples are research on soft actuators \cite{kim2019review,yao2024multimodal,xin2024role}
and soft robotics \cite{whitesides2018soft,trivedi2008soft,yao2024multimodal,abhyankar2024development}. 
Often, actuation by external stimuli in a contactless way is beneficial or mandatory. Magnetic fields stand out in this regard \cite{rehman2024magnetic}. They offer remote control and, mostly, they are biologically harmless and do not pose any risk to health. 

Both aspects are addressed and combined by magnetic elastomers. Microscopic magnetic or magnetizable particles of up to about a hundred micrometers in size are enclosed by a soft, deformable, polymeric carrier matrix. When applying external magnetic fields, the materials respond. Specifically, they deform, which makes them promising candidates, for instance, for soft actuators \cite{filipcsei2007magnetic,an2003actuating,zrinyi1996deformation,ilg2013stimuli,collin2003frozen,raikher2008numerical,fuhrer2009crosslinking,schmauch2017chained,hines2017soft} or magnetic valves \cite{bose2012soft,yoo2002design}. 
Additionally, they change their overall mechanical behavior, which is referred to as magnetorheological effect. Their stiffness, quantified in terms of their elastic Young or shear modulus, can increase by an order of magnitude \cite{filipcsei2007magnetic,pessot2014structural,wood2011modeling,ivaneyko2012effects,evans2012highly,han2013field,borin2013tuning,zubarev2019rheological,volkova2017motion,roghani2025magnetically,bastola2020recent}.
Likewise, their dissipative behavior is affected, suggesting them as tunable damping devices and vibration absorbers \cite{deng2006development,liao2012development,molchanov2014viscoelastic,sun2008study,li2013highly}. 

From early attempts of structuring the arrangement of the magnetic particles inside the elastomers, we know that controlled particle positioning can significantly enhance the material properties. Specifically, during the manufacturing process, strong homogeneous magnetic fields were applied to fluid suspensions of magnetic particles in reactive solutions. Anisotropic particle aggregates formed \cite{li2014state,jolly1996magnetoviscoelastic,bose2007viscoelastic,borbath2012xmuct,gunther2011x,danas2012experiments,borin2020targeted,collin2003frozen,coquelle2005magnetostriction}.
Then, polymerization of these structured suspensions and chemical crosslinking to form an elastomeric matrix was triggered. This process permanently locked in the anisotropic particle arrangements. When the structuring external magnetic field is turned off, the particle aggregates are maintained by the confining elastic carrier matrix. As a consequence of the established structuring, the magnetorheological effect can be enhanced substantially \cite{bellan2002field,bose2007viscoelastic}. Meanwhile, modern patterning techniques such as 3D printing \cite{khalid20243d,qi20203d,zhang20214d,bayaniahangar20213,kim2018printing, dohmen2020field,bastola2020dot,cao20213d}, sequential photopolymerization \cite{kim2011programming}, templated particle positioning \cite{puljiz2016forces,zhang2008analysis}, structuring by magnetic fields \cite{martin2006magnetostriction}, acoustic holography \cite{peerfischer}, and wax-cast molding \cite{forster2013patterning} 
have been presented. Through their advent and further development, controlled positioning of magnetizable inclusions in elastomeric carrier matrices comes into reach. 

Therefore, we recently addressed the question of optimizing the positioning of magnetizable inclusions in the elastic carrier matrix to maximize the requested effects \cite{fischer2024opt}. That is, we identified specific structural arrangements that most strongly enhance the magnetically induced relative changes in elastic moduli, that is, the magnetorheological effect. For this purpose, we developed a computational approach that maximizes the requested behavior as a function of the positioning of individual magnetizable sites \cite{fischer2024opt}. These magnetizable sites can consist of individual particles, magnetic droplets during 3D printing, or isotropic particle clusters. We imposed a minimum distance of these magnetic sites between each other and from the surface of the system. These gaps are filled by nonmagnetic elastomer, which we assume reasonable in view of 3D printing. The identified spatial arrangements of magnetizable inclusions substantially enhance the magnetorheological effect when compared to isotropic positioning. 

Here, we proceed an additional significant step forward in our recipe of enlarging the magnetorheological effect. 
We understand by the magnetorheological effect generally the relative change in elastic modulus that can maximally be achieved by the action of external magnetic fields to a given system. 
Specifically, we point out a way of further increasing it by approximately a factor of two by doubling it in magnitude.
Considering the relevance of contactless magnetic tuning of soft mechanical properties, we expect this basic idea to bear significant technological potential. 

Key to our concept is to work with two perpendicular magnetic fields. We not only consider magnetically induced mechanical hardening but also mechanical softening. 
In our recent work, we optimized the internal structural arrangement for maximized increase in mechanical stiffness \cite{fischer2024opt}. Yet, we also identified maximized decrease in mechanical stiffness \cite{fischer2024opt}. This ``negative'' magnetorheological effect has hardly been addressed so far \cite{han2013field,yang2014novel}. 
The basis for doubling the magnitude of the magnetorheological effect now is to combine the two antagonistic types of behavior. We optimize the arrangement of the magnetizable inclusions to facilitate both, hardening and softening, in the same system. Whether then, in action, the system hardens or softens is selected by the magnetic field direction. That is, we identify optimized positioning of the magnetizable sites in a way that maximizes mechanical hardening when applying the magnetic field in one direction. On the contrary, the structure implies maximized softening when applying the magnetic field in a perpendicular direction. Switching between these two mechanical antipodes by redirecting the magnetic field implies an approximately doubled magnetorheological effect, when taking the softened state as a reference and switching to the hardened state. Instead, conventional previous considerations addressed hardening under magnetization, starting from the nonmagnetized state as a reference. 

\begin{figure}
	\includegraphics[width=\linewidth,trim={0cm 0.2cm 0cm 0cm}, clip]{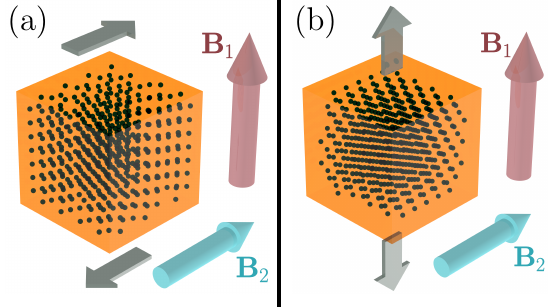}
	\caption{Illustration of the considered geometries. We consider a cubical elastic system (orange) with $N$ magnetizable inclusions (dark dots). It is subject to an externally imposed mechanical deformation as indicated by the flat arrows. Both (a) shear and (b) uniaxial elongation are addressed. An external magnetic field is either applied in a direction as indicated by the cylindrical red arrow $\mathbf{B}_1$, or perpendicular to that, as marked by the cylindrical turquoise arrow $\mathbf{B}_2$. The system is optimized concerning the positions of the magnetizable inclusions for elastic hardening/stiffening under magnetic field application along one direction and elastic softening under magnetic field application along the other direction. Switching between the two magnetic field directions then implies an approximately doubled magnetorheological effect compared to the nonmagnetized situation as a state of reference.}
	\label{fig:setup}
\end{figure}

To demonstrate the doubled magnetorheological effect, we proceed as follows. All our considerations are theoretical in nature and based on computational evaluations, in the spirit of computational materials design.
We focus on cubical systems. 
They consist of an incompressible, linearly elastic carrier medium with embedded magnetizable inclusions of vanishing magnetic remanence. 
Our task is to sort the magnetizable inclusions into the elastic cube in a way that maximizes the overall elastic  modulus of the system when magnetized in one direction. At the same time, the overall elastic  modulus shall be minimized when applying the magnetic field in a perpendicular direction. Thus, the first magnetic field direction implies maximized elastic hardening/stiffening. The second, perpendicular field direction leads to maximized elastic softening. 

To approach this task, we start from our previously developed computational scheme \cite{fischer2024opt}. It is based on simulated annealing \cite{MC,kirkpatrick1983optimization,vanderbilt1984monte} with adaptive cooling rates \cite{karabin2020simulated}. 
There, for a given number of magnetizable inclusions $N$, we optimized their positions to achieve \textit{either} maximized elastic hardening/stiffening \textit{or} maximized elastic softening under application of {only one} magnetic field. The reference state to quantify the hardening/stiffening and softening was, as conventional until now, the nonmagnetized state of the system.
Moreover, we assumed all inclusions to be identical and magnetized to saturation when the external magnetic field is applied. We considered affine deformations of the elastic material. 
The output was given by spatial arrangements of the magnetizable inclusions in the cube \cite{fischer2024opt}. 

For our present purpose of demonstrating the doubled magnetorheological effect, we have modified our computational scheme accordingly. 
We now optimize the spatial arrangement of the magnetizable inclusions such that the system shows a maximized difference in elastic  modulus when an applied external magnetic field is switched between two perpendicular directions.

To illustrate the doubled magnetorheological effect, we focus on two geometries, see Fig.~\ref{fig:setup}. First, we consider imposed shear deformations of the cubical system, see Fig.~\ref{fig:setup}(a). The two perpendicular magnetic field directions $\mathbf{B}_1$ and $\mathbf{B}_2$ are each normal to a pair of surfaces of the cube. Simultaneously, the shear planes contain both perpendicular magnetic field directions. Second, we address uniaxial stretching, as illustrated in Fig.~\ref{fig:setup}(b). Lateral contraction follows from the assumed incompressibility of the material. Here, one magnetic field direction $\mathbf{B}_1$ is along the stretching axis. The other magnetic field orientation $\mathbf{B}_2$ is perpendicular to the stretching axis. Each direction is normal to a pair of surfaces of the cube.

%\section{Results and discussion}
\begin{figure}
	\includegraphics[width=\linewidth]{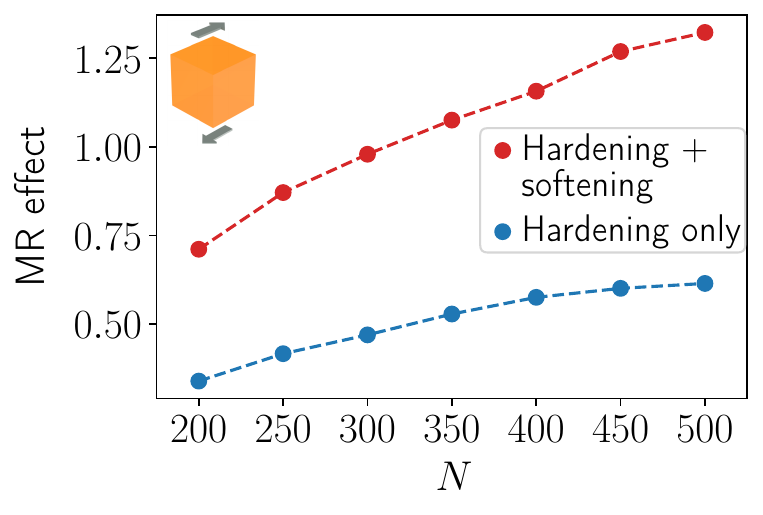}
	\caption{Magnetically induced relative change in elastic shear  modulus, that is, magnitude of the magnetorheological effect (MR effect) under imposed shear deformation, see the geometry in Fig.~\ref{fig:setup}(a), as a function of the number of magnetizable inclusions $N$.
		The blue curve includes our previously obtained data points that were obtained by an optimization strategy for the positions of the magnetizable inclusions \cite{fischer2024opt}. Yet, only mechanical hardening/stiffening by application of one magnetic field was considered relative to the nonmagnetized state of the system. Instead, we now switch between two perpendicular magnetic field directions, see Fig.~\ref{fig:setup}(a). One field triggers maximized mechanical hardening/stiffening, the other one induces maximized mechanical softening. The resulting relative change in elastic modulus is more than doubled in magnitude (red curve).
	}
	\label{fig:results1}
\end{figure}

We start with the case of externally imposed shear deformations, as displayed in Fig.~\ref{fig:setup}(a). 
To achieve the maximized overall change in shear modulus, we optimize the spatial configuration of the magnetizable inclusions to maximize elastic hardening/stiffening under application of the magnetic field $\mathbf{B}_1$, while simultaneously optimizing for maximized elastic softening under application of the perpendicular magnetic field $\mathbf{B}_2$. 
The procedure is performed for different fixed numbers of magnetic inclusions $N$. Figure~\ref{fig:results1} displays by the red curve the resulting relative change in elastic shear  modulus when switching between the two perpendicular magnetic field directions $\mathbf{B}_1$ and $\mathbf{B}_2$. As expected, the magnetically induced relative change in shear modulus increases with increasing number of inclusions $N$.
For comparison, we also include our previous results \cite{fischer2024opt} for maximized elastic hardening under application of the magnetic field $\mathbf{B}_1$, but with the reference state set as the nonmagnetized system, see the blue curve. Choosing the nonmagnetized state as a reference has been conventional until now. 

Our optimization strategy now allows to set as a reference state the perpendicularly magnetized, mechanically softened state of the system.   
We can infer that, as a consequence of the perpendicular switching of the magnetic field in combination with the bidirectional optimization strategy, we have more than doubled the magnetorheological effect, comparing the red and the blue curves.
For the highest considered number of inclusions $N=500$, the magnetically induced relative change in elastic shear  modulus (red curve) is approximately of a factor $2.15$ larger
when compared to the previously observed, already  maximized effect working with only one magnetic field direction (blue curve).

\begin{figure}
	\includegraphics[width=\linewidth]{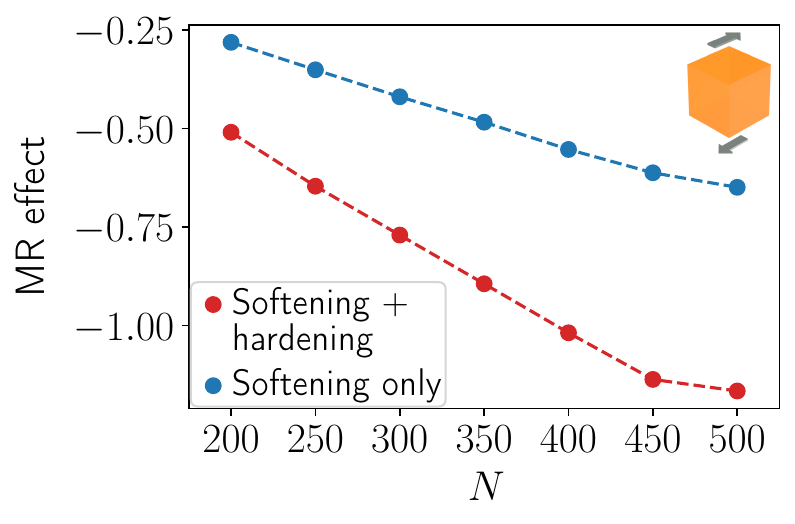}
	\caption{Similar to Fig.~\ref{fig:results1}, but now associating maximized mechanical softening with the magnetic field direction $\mathbf{B}_1$ and maximized elastic hardening/stiffening with the perpendicular magnetic field direction $\mathbf{B}_2$. The relative change in elastic shear  modulus for mechanical softening is negative, see the sign on the vertical axis. Again, the magnetically induced relative change in shear modulus when working with two perpendicular magnetic field directions (red curve) is approximately doubled when compared to the previous optimization scheme \cite{fischer2024opt} of using only one magnetic field direction for mechanical softening relative to the nonmagnetized state (blue curve).
	}
	\label{fig:results2}
\end{figure}

For completeness, we also consider the opposite switching scenario induced by the perpendicular magnetic fields. That is, we optimize the spatial configuration of the magnetizable inclusions for maximized elastic softening when the magnetic field $\mathbf{B}_1$ is applied. Simultaneously, we optimize for maximized elastic hardening/stiffening under application of the magnetic field $\mathbf{B}_2$. The results are again convincing, see Fig.~\ref{fig:results2}. Here, the magnitude of the induced relative change in elastic modulus is a bit less then doubled (red curve) when comparing to the previously optimized scenario that only considers magnetically induced mechanical softening relative to the nonmagnetized state of the system (blue curve). 
Since mechanical softening is associated with a negative change in the magnitude of the elastic shear modulus, this scenario was termed negative magnetorheological effect \cite{han2013field,yang2014novel}.
For the highest considered number of magnetizable inclusions $N=500$, the relative change in elastic shear  modulus in Fig.~\ref{fig:results2} is approximately by a factor of $1.80$ larger (red curve) than the previously maximized value (blue curve).

\begin{figure}
	\includegraphics[width=\linewidth]{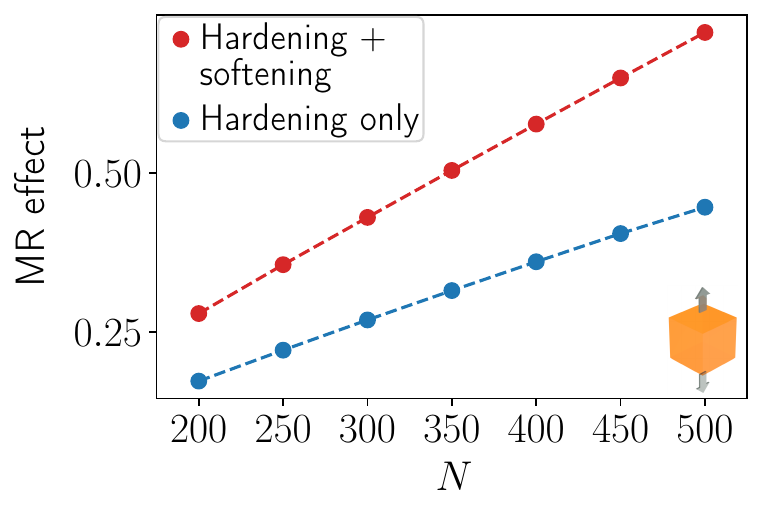}
	\caption{Analogous to Fig.~\ref{fig:results1}, but now for the relative change in elastic  Young/stretching modulus. In other words, we display the relative change in elastic  Young/stretching modulus when switching from $\mathbf{B}_2$ in Fig.~\ref{fig:setup}(b) to $\mathbf{B}_1$ (red curve). We also include our previous results obtained by maximizing elastic hardening/stiffening relative to the nonmagnetized state of reference when turning on the magnetic field $\mathbf{B}_1$ from zero (blue curve) \cite{fischer2024opt}.
	}
	\label{fig:results3}
\end{figure}

Additionally, we consider imposed uniaxial stretching as indicated in Fig.~\ref{fig:setup}(b). 
Here, we evaluate the magnetically induced relative change in elastic  Young/stretching modulus. We simultaneously optimize the spatial arrangement of the magnetizable inclusions for maximized mechanical hardening/stiffening under application of the magnetic field $\mathbf{B}_1$ and for maximized mechanical softening under application of the perpendicular magnetic field $\mathbf{B}_2$. 
Corresponding results are displayed by the red curve in Fig.~\ref{fig:results3}. 
Analogously to the elastic shear modulus, the magnetically induced relative change in elastic Young modulus is substantially enlarged (red curve). It is compared to previous results for optimized configurations of maximized elastic hardening relative to the nonmagnetized state (blue curve) \cite{fischer2024opt}. For the number of magnetizable inclusions $N=500$, we here observe an increase in the relative change of elastic Young modulus by a factor of approximately $1.62$. % for $N=500$ where the combined MR effect reaches a value of approximately $72\%$.

\begin{figure}
	\includegraphics[width=\linewidth]{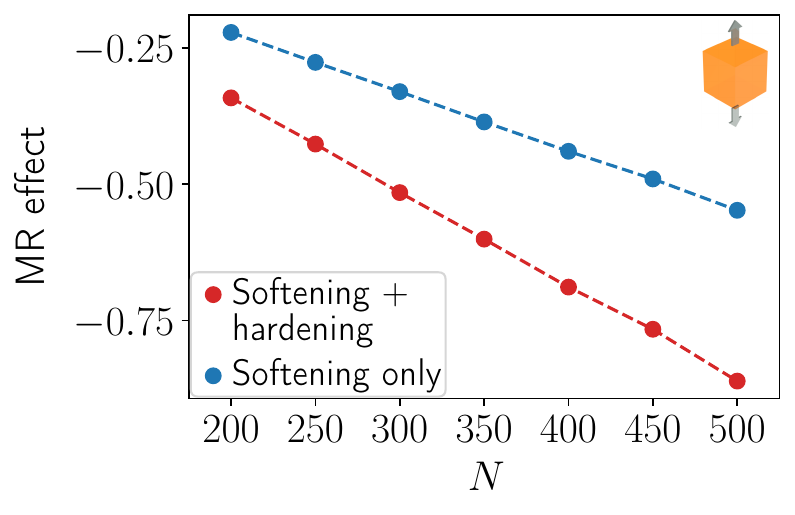}
	\caption{
		Similar to Fig.~\ref{fig:results3} for imposed uniaxial stretching, but now inducing maximized mechanical softening by the magnetic field $\mathbf{B}_1$ and maximized mechanical hardening/stiffening for the perpendicular magnetic field $\mathbf{B}_2$. Considering maximized magnetically induced softening, the sign of the relative change in elastic Young/stretching modulus is negative. We reveal a significantly enhanced relative reduction in elastic Young/stretching modulus when switching between the two perpendicular magnetic fields $\mathbf{B}_1$ and $\mathbf{B}_2$ for the correspondingly optimized structure (red curve). Previous data for maximized magnetically induced softening for only turning on $\mathbf{B}_1$ from zero are included for reference (blue curve) \cite{fischer2024opt}.}
	\label{fig:results4}
\end{figure}

Again, for completeness, we address the switched situation of optimizing for maximized softening under application of the magnetic field $\mathbf{B}_1$, while simultaneously optimizing for maximized hardening/stiffening under application of the perpendicular magnetic field $\mathbf{B}_2$.
Figure~\ref{fig:results4} displays the corresponding results for the resulting maximized relative reduction in elastic  Young/stretching modulus when switching from $\mathbf{B}_2$ to $\mathbf{B}_1$. Once more, our strategy of switching between two perpendicular external magnetic fields for correspondingly optimized spatial arrangements of the magnetizable inclusions (red curve) leads to significantly enhanced relative changes in elastic Young modulus. We compare to the situation of an optimized configuration for maximized mechanical softening relative to a nonmagnetized state of reference (blue curve) \cite{fischer2024opt}. Due to the considered mechanical softening, the sign of the relative change in elastic modulus in Fig.~\ref{fig:results4} is again negative. The maximal relative change in Young modulus for the number of magnetizable inclusions $N=500$ was about a factor of $1.57$ larger (red curve) when compared to the previous results of optimizing the structure for switching from the nonmagnetized state (blue curve) \cite{fischer2024opt}.

For the last geometry, we include as a supporting information additional data on the spatial arrangement of the magnetizable inclusions in the cube. In this case, we can nicely infer the influence of the second, perpendicular magnetic field when optimizing the structural arrangement for the alternating action of the two perpendicular magnetic fields. Considering only $\mathbf{B}_1$ along the stretching axis, the resulting transversal cubical symmetry in the spatial arrangement of the inclusions in the plane perpendicular to the stretching axis is approximately maintained. However, simultaneously considering the alternatively applied perpendicular magnetic field $\mathbf{B}_2$ in the optimization scheme breaks this transversal cubical symmetry, which is reflected by the spatial organization of the magnetizable inclusions.

%\section{Conclusions}
%\label{sec:conclusions}

In conclusion, we here outline a path of designing magnetorheological materials that feature a substantially enhanced magnetically induced relative change in mechanical properties. Our strategy is based on positioning the magnetizable inclusions in a soft, elastic carrier medium in a way that is most beneficial for the desired purpose. Already in a previous investigation, we optimized the spatial arrangement of the magnetizable inclusions in a corresponding manner \cite{fischer2024opt}. Yet, there, we still considered the nonmagnetized states of the systems as the states of reference to quantify the magnetically induced relative change in mechanical properties. Generally, this has been the conventional point of reference so far. 

Equipped with our modified computational formalism, we here employ switching between two perpendicular magnetic fields. The spatial arrangement of the magnetizable inclusions is optimized in a way that one magnetic field direction induces maximized elastic hardening. The other magnetic field direction implies maximized elastic softening. In combination, switching between these two perpendicular magnetic fields, the magnitude of induced relative change in elastic moduli is approximately doubled when compared to the situation that employs only one magnetic field direction. 

Not necessarily are we thinking of rigid magnetizable inclusions, but also droplets of magnetic fluid could be considered in reality, for instance, during 3D printing processes \cite{qi20203d, bastola2020dot}. This should be helpful when our suggestions of materials design are transferred to reality. Soft magnetic elastomers have a large potential when it comes to magnetically tuning mechanical properties. Promoting an approximate doubling in magnitude of these effects should raise enhanced technological interest in these materials.

\subsection*{Supporting information}
%Supporting information: 
Further information is included on the optimized configuration for uniaxial stretching that illustrates symmetry breaking caused by the additional optimization with respect to a perpendicular magnetic field (PDF).\\

\subsection*{Acknowledgments}
We thank Peter Harrowell for a stimulating discussion at the beginning of this work and for valuable feedback on the manuscript. Moreover, we acknowledge support by the Deutsche Forschungsgemeinschaft (German Research Foundation, DFG) through the Research Unit FOR 5599 on structured magnetic elastomers, project no.\ 511114185, via DFG grant reference nos.\ ME 3571/10-1 and ME 3571/11-1, and through the Heisenberg Grant, project no.\ 413993216, via DFG grant reference no.\ ME 3571/4-1.

\bibliography{literature}

\end{document}

	%%%
	
	%%%%%General information
	
	%%% Title 
	\title{
		Supporting Information \\
		``Doubling the magnetorheological effect of magnetic elastomers''
	}
	
	%%% Authors
	\author{Lukas Fischer}
	%\email{lukas.fischer@ovgu.de}
	\affiliation{Institut f{\"u}r Physik, Otto-von-Guericke-Universit{\"a}t Magdeburg, Universit{\"a}tsplatz 2,  39106 Magdeburg, Germany}
	\author{Andreas M. Menzel}
	%\email{a.menzel@ovgu.de}
	\affiliation{Institut f{\"u}r Physik, Otto-von-Guericke-Universit{\"a}t Magdeburg, Universit{\"a}tsplatz 2,  39106 Magdeburg, Germany}
	
	%%% Date
	\date{\today}

	%%% Abstract
	\begin{abstract}
		In the main text, we describe how optimized arrangements of magnetizable inclusions lead to maximized magnetorheological effects under switching of external magnetic fields between mutually perpendicular orientations. Here, we include additional information about these optimized configurations for the case of uniaxial stretching. The twofold optimization reduces the symmetry of the resulting structure. 
	\end{abstract}
	
	%%% Maketitle
	\maketitle
    We focus for illustration on the geometry of uniaxial stretching. Applying external magnetic fields affects the associated Young modulus of the system. 

    Specifically, we first optimize the configuration for maximized softening when applying an external magnetic field $\mathbf{B}_1$ along the stretching axis, see Fig.~1(b) in the main text. 
\begin{figure}
	\includegraphics[width=\linewidth]{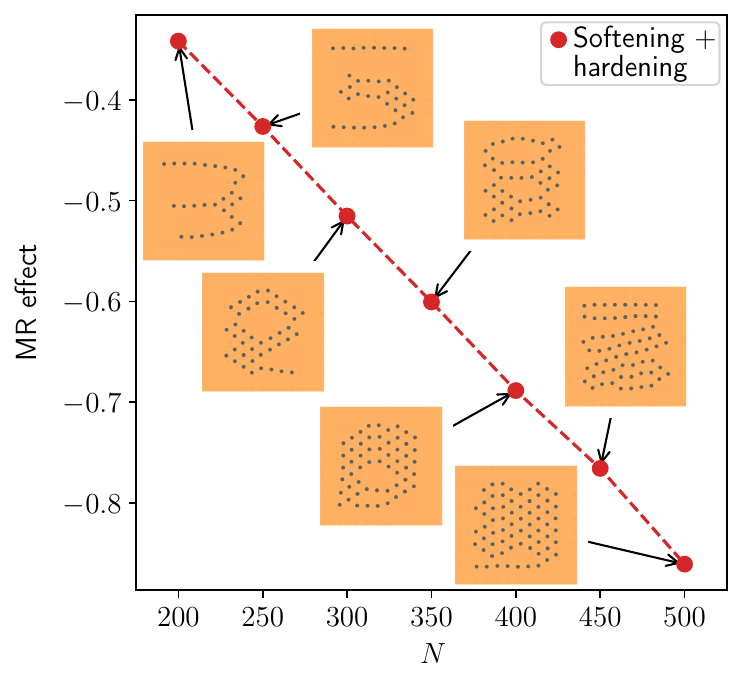}
    %\vspace*{-0.8cm}
    \caption{Same as Fig.~4 of the main article, indicating the increase in magnitude of the magnetorheological (MR) effect  with the number of inclusions $N$. Here, top views of the resulting configurations for doubling the magnetorheological effect are added. Each dot in the snapshots marks the location of a chain-like aggregate oriented along the stretching direction as viewed from the top, thus pointing towards the reader. We clearly see how the additional optimization of the arrangements with respect to a magnetic field perpendicular to the stretching direction, here oriented horizontally, leads to additional anisotropy of the structures, specifically at lower values of $N$.
    }
	\label{fig:s1}
\end{figure}
    In our previous work \cite{fischer2024opt}, we found that the associated optimized configurations
    mainly consist of chain-like aggregates. They are oriented along the stretching axis and magnetic field direction. 
    Overall, within the transverse plane, we find approximate isotropy of the configurations of the inclusions (see also Fig.~7B in Ref.~\onlinecite{fischer2024opt}), to some degree broken by the cubical geometry.
    
    Now, we perform additional optimization for maximized hardening when the external magnetic field $\mathbf{B}_2$ is alternately applied in a direction perpendicular to the stretching axis and $\mathbf{B}_1$. Obviously, such an external magnetic field breaks rotational symmetry around the stretching axis and thus in the transversal plane.
    Indeed, we observe additional, anisotropic, chain-like organization of the inclusions in the direction perpendicular to the stretching axis, along $\mbf{B}_2$, see Fig.~\ref{fig:s1}. The effect is particularly obvious for lower numbers of inclusions $N$. 
	\bibliography{literature}
	\newpage